\documentclass{aastex631}

\begin{document}

\title{The Spectrum of C/2023 A3 Indicates A Depleted Composition}

\correspondingauthor{Shihao Wang}
\email{wangshihao99@126.com}








\author[0009-0007-3777-1561]{Yunyi Tang}
\affiliation{Department of Astronomy, Tsinghua University, Beijing 100084, People's Republic of China}
\affiliation{Tsinghua University Student Astronomy Society, Tsinghua University, Beijing 100084, People's Republic of China}

\author[0009-0001-8606-2813]{Shihao Wang}
\affiliation{Department of Chemistry, Tsinghua University, Beijing 100084, People's Republic of China}
\affiliation{Tsinghua University Student Astronomy Society, Tsinghua University, Beijing 100084, People's Republic of China}

\author[0009-0009-7223-2285]{ZiXuan Lin}
\affiliation{Academy of Art and Design, Tsinghua University, Beijing 100084, People's Republic of China}
\affiliation{Tsinghua University Student Astronomy Society, Tsinghua University, Beijing 100084, People's Republic of China}

\author[0009-0008-6685-921X]{Xiaorui Yang}
\affiliation{Department of Astronomy, Tsinghua University, Beijing 100084, People's Republic of China}
\affiliation{Tsinghua University Student Astronomy Society, Tsinghua University, Beijing 100084, People's Republic of China}

\author[0009-0002-9243-0819]{Xinyang Zhang}
\affiliation{Weixian College, Tsinghua University, Beijing 100084, People's Republic of China}
\affiliation{Tsinghua University Student Astronomy Society, Tsinghua University, Beijing 100084, People's Republic of China}

\author[0009-0006-5531-536X]{Songyu Jia}
\affiliation{Press of Aerospace Knowledge, Beijing 100083, People's Republic of China}

\author[0000-0002-6937-9034]{Sharon X.~Wang}
\affiliation{Department of Astronomy, Tsinghua University, Beijing 100084, People's Republic of China}



\begin{abstract}

We report a spectroscopic observation of comet C/2023 A3 using an 80 mm apochromatic (apo) refractor equipped with a custom-built spectrometer with a resolution of R$\sim$2,500 on the night of 4 October 2024. Sodium D lines were detected prominently, while no other emission lines, particularly carbon-bearing species, were observed, which suggests that comet C/2023 A3 may be carbon-depleted. The mobility and flexibility of our observational setup highlight the value of amateur telescopes in observing low-altitude targets like C/2023 A3 as a complement to professional facilities.

\end{abstract}



\section{Introduction} 

C/2023 A3, also known as Tsuchinshan-ATLAS, is a long-period comet discovered independently by the Purple Mountain Observatory in China on 9 January 2023 and ATLAS South Africa on 22 February 2023 \citep{Ye2023}. It reached its perihelion on 27 September 2024, at a distance of 0.391 AU \citep{Sato2024}, making it one of the brightest comets visible to the naked eye in 2024, attracting significant interest from professional and amateur astronomers.

Comets are scientifically valuable as they are remnants from the early solar system, preserving information about primordial conditions \citep[e.g.,][]{BRASSER201340,2015A&A...583A..43M}. Depleted comets, characterized by a deficiency in carbon-chain molecules such as C${2}$, C${3}$, and CN in their spectra, represent only 9\% of observed comets \citep{COCHRAN2012144}. Studying these comets can provide insights into the environmental conditions of the early solar nebula and their distinct evolutionary processes.

Due to limitations in observing time and telescope availability for many professional facilities, amateur astronomers can play an important role in collecting timely data. Their flexibility allows for observations under conditions that may not be practical for larger, fixed installations. In this study, we utilized an 80 mm apochromatic refractor telescope equipped with a custom-built spectrometer to observe comet C/2023 A3, aiming to contribute complementary spectroscopic data on its chemical composition.

\section{OBSERVATIONS AND DATA PROCESSING} 

\begin{figure}[ht!] 
\centering 
\resizebox{\textwidth}{!}{\includegraphics{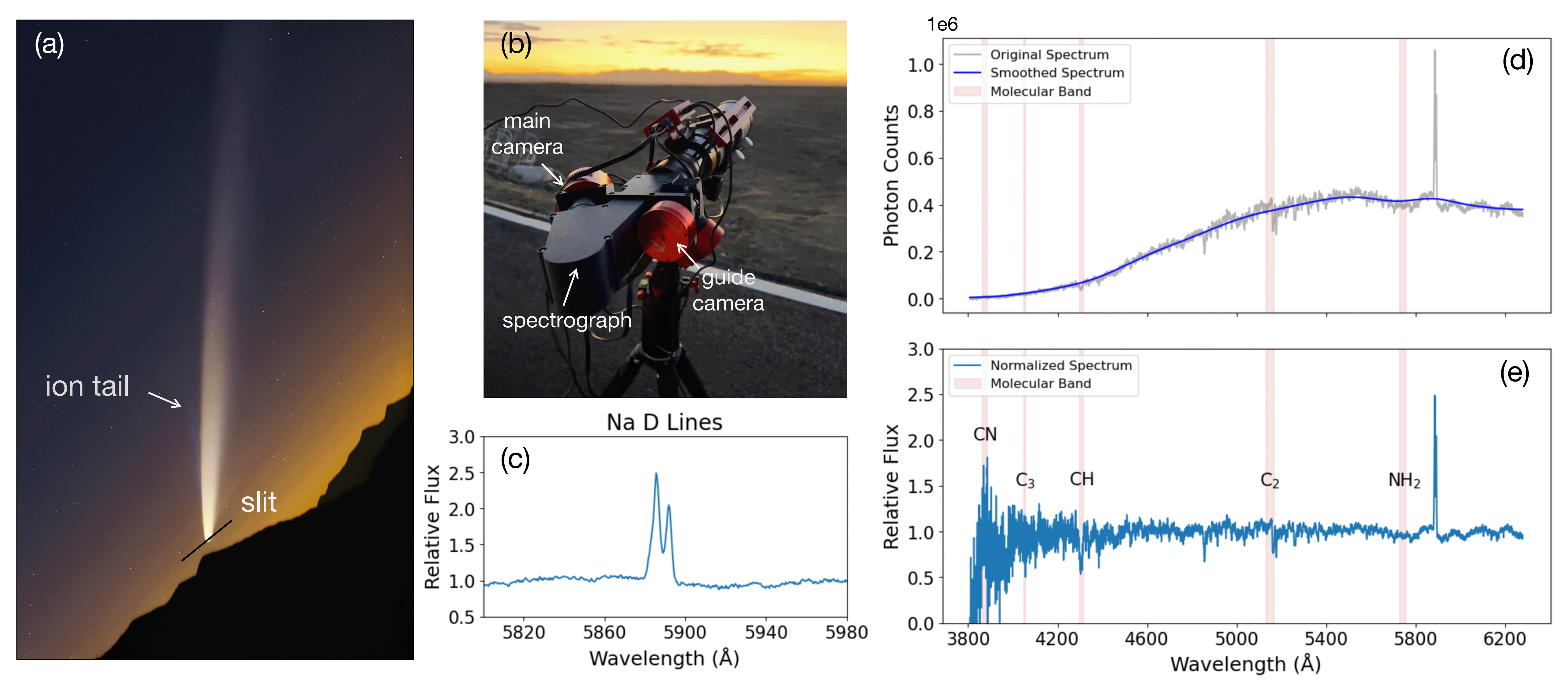}} 
\caption{(a) Image of C/2023 A3 showing the slit position. The faint blue ion tail is visible to the left of the bright dust tail (with a significantly enhanced contrast). (b) Image of our equipment. (c) The normalized spectrum highlighting the sodium D lines. (d) The extracted spectrum of C/2023 A3 shows clear sodium doublet at 5890 \textup{\AA} and 5896 \textup{\AA}. The wavelength ranges of some common emission lines are marked. (e) The normalized spectrum.} 
\label{fig:combined} 
\end{figure}

Observations of comet C/2023 A3 were conducted using a SkyRover 80 mm f/6 apochromatic refractor telescope, coupled with a custom-built spectrometer featuring a 12 mm long and 14 $\mu$m wide slit, and a 600 grooves/mm reflective grating. Our camera was a ZWO asi183mm with a CMOS sensor of 5496×3672 pixels and 2.4 $\mu$m per pixel. Our telescope and instrument setups are shown in the photo in panel (b) of Figure~\ref{fig:combined}. The observations took place at Lenghu Observatory on Saishiteng Mountain, Qinghai Province \citep{Deng2021}, on 4 October 2024, when the comet was visible before sunrise.

The operating range of the spectrograph is 300-860 nm, and we adjusted our wavelength range to be 380--620 nm, achieving a resolution of 2500 at 577 nm with a pixel scale of $\sim$0.48 \AA\ per pixel. The spectrograph was equipped with a guide camera to capture the slit plate's reflected image, automatically guiding on the comet’s nucleus. An approximate position of the slit with respect to the comet is illustrated in panel (a) of Figure~\ref{fig:combined}. We took 43 frames with 30-second exposures, totaling 1290 seconds, from approximately 6:44 am to 7:07 am with altitudes between 1.0$^{\circ}$ and 5.1$^{\circ}$.

Raw data reduction and spectral extraction were performed using a customized Python pipeline. We conducted bias and dark subtraction, along with calibration spectral images of the sky and Procyon. The reduced Procyon spectral image was used to derive the rotation angle to align the dispersion direction of the slit with the pixel rows. We then further rectified the images in the cross-dispersion direction: each row was shifted until it aligned with a fiducial row in the reduced and stacked sky spectral image. We then applied these shifts to all reduced spectral images of the comet. Flat fielding was performed in each column in the cross-dispersion direction after normalizing each row of the rectified sky spectral image.

For the reduced comet spectral images, we defined a wide extraction window of 400 pixels to include most light from the comet and estimated the sky background by averaging 400 pixels both above and below this window. We then stacked the 42 sky-subtracted spectral images to extract the comet's spectrum within a narrower, high-SNR window of 52 pixels. The extracted spectrum is shown in panel (d) of Figure~\ref{fig:combined}.

Wavelength calibration was achieved by fitting known absorption lines in Procyon’s spectrum and apllied to the comet's spectrum. We applied Gaussian filtering to estimate the continuum, resulting in the normalized spectrum shown in panel (e) of Figure \ref{fig:combined}. The median SNR of the extracted spectrum of C/2023 A3 is approximately 600 per extracted pixel or  $>$1400 per resolution element ($\sim$5 pixels or $\sim$2.3 \AA).

\section{RESULTS AND DISCUSSION} 

We detected sodium D lines at 5890 \textup{\AA} and 5896 \textup{\AA} (panel c of Figure~\ref{fig:combined}), indicating the presence of neutral sodium in the comet’s nucleus. However, we were unable to detect other expected emission lines, such as CN, CH, C${2}$, and C${3}$, compared to comet C/2020 F3 (NEOWISE; \citealt{2021A&A...656A.160C}). The spectrum has low SNR near the CN line, making definitive conclusions difficult. The CH line coincides with a solar absorption feature, complicating the detection.

This lack of carbon-related emissions suggests that comet C/2023 A3 may be carbon-depleted. However, data processing limitations and high atmospheric extinction could also influence these results. Nonetheless, our high-SNR spectrum rules out high-intensity emissions from carbon molecules commonly observed in other comets \citep{COCHRAN2012144}.

The absence of carbon emissions aligns with the weak ion tail of C/2023 A3, as illustrated in panel (a) of Figure~\ref{fig:combined}. To capture the ion tail, we used the SkyRover 70SA Telescope and Canon 6D2 DSLR camera, enhancing the B-channel brightness significantly, resulting in the faint blue ion tail observed.

Our observations demonstrate the capabilities of amateur equipment in complementing professional telescopes for low-altitude targets. The lightweight 80 mm refractor effectively tracked the comet at very low altitudes ($\approx$1$^{\circ}$), showcasing the flexibility of amateur setups.

In the future, we plan to continue monitoring the spectrum of C/2023 A3 to characterize its temporal variation. We will also enhance our data reduction and calibration processes, including more thorough flat fielding and solar spectrum subtraction.

\section*{Acknowledgments} 
\begin{acknowledgments} 
We acknowledge the Tsinghua educational funds, the Tsinghua Dushi Funds (grant number 53121000124), and the Tsinghua SRT Project Funds (grant number 53411000123). We thank Mr. Bin Dong, the local administrator of Lenghu Town, for his assistance during observations. This research utilized the Astrophysics Data System, funded by NASA under Cooperative Agreement 80NSSC21M00561. 
\end{acknowledgments}

\software{astropy \citep{2013A&A...558A..33A,2018AJ....156..123A}}

%




\bibliography{sample631}{}
\bibliographystyle{aasjournal}



\end{document}